\documentstyle[11pt,amssymb,amsfonts]{article}

\begin{document}
\newcommand{\p}{\parallel }
\makeatletter \makeatother
\newtheorem{th}{Theorem}[section]
\newtheorem{lem}{Lemma}[section]
\newtheorem{de}{Definition}[section]
\newtheorem{rem}{Remark}[section]
\newtheorem{cor}{Corollary}[section]
\renewcommand{\theequation}{\thesection.\arabic {equation}}

\title{{\bf The spectral action for sub-Dirac operators}
}
\author{Yong Wang \\}

\date{}
\maketitle

\begin{abstract} In this paper, for foliations with spin leaves,
we compute the spectral action for sub-Dirac operators.\\

\end{abstract}

\section{Introduction}
    \quad Connes'spectral action principle ([Co]) in noncommutative
    geometry states that the physical action depends only on the
    spectrum. We assume that space-time is a
    product of a continuous manifold and a finite space. The
    spectral action is defined as the trace of an arbitrary function
    of the Dirac operator for the bosonic part and a Dirac type
    action of the fermionic part including all their interactions.
    In [CC1], Chamseddine and Connes computed the Spectral action for
    Dirac operators on spin manifolds and the Chamseddine-Connes
    spectral action comprises the Einsiein-Hilbert action of general
    relativity and the bosonic part of the action of the standard
    model of particle physics. In [HPS], Hanisch, Pf\"{a}ffle and
    Stephan derived a formula for the gravitional part of the
    spectral action for Dirac operators on $4$-dimensional spin
    manifolds with totally anti-symmetric torsion. They also deduced
    the Lagrangian for the Standard Model of particle Physics in the
    presence of torsion from the Chamseddine-Connes
    spectral action. In [CC2], Chamseddine and Connes studied the
    spectral action for spin manifolds with boundary and generalized
    this action to noncommutative spaces which are products of a
    spin manifold and a finite space. In [EILS],[ILS], the spectral
    actions for the noncommutative torus and $SU_q(2)$ are computed
    explicitly.\\
      \indent In this paper, we consider a compact foliation $M$ with spin
      leaves. We don't assume that $M$ is spin, so we have no Dirac
      operators on $M$, then we can not derive the physical action
      from the Chamseddine-Connes spectral action for Dirac operators.
      In [LZ], in order to prove the Connes' vanishing theorem for
      foliations with spin
      leaves, Liu and Zhang introduced sub-Dirac operators
      instead of Dirac operators. The sub-Dirac operator is a first
      order formally self adjoint elliptic differential operator. So we
      have a commutative spectral triple and we compute the spectral
      action for sub-Dirac operators.\\
      \indent This paper is organized as follows: In Section 2, we review
      the sub-Dirac operator and compute the spectral
      action for sub-Dirac operators.
      In Section 3, we compute the spectral
      action for sub-Dirac operators for the Standard Model.In Section 4, we compute the spectral
      action for sub-Dirac operators for foliations with boundary.\\

\section{The spectral
      action for sub-Dirac operators }

 \quad Let $(M,F)$ be a closed foliation and $g^F$ be a metric on
 $F$. Let $g^{TM}$ be a metric on $TM$ which restricted to $g^F$ on
 $F$. Let $F^\perp$ be the orthogonal complement of $F$ in $TM$ with
 respect to $g^{TM}$. Then we have the following orthogonal
 splitting,
 $$TM=F\oplus F^\perp;~~g^{TM}=g^F\oplus g^{F^\perp},\eqno(2.1)$$
 where $g^{F^\perp}$ is the restriction of $g^{TM}$ to $F^\perp$.
 Let $P,P^\perp$ be the orthogonal projection from $TM$ to $F$,$F^\perp$
 respectively. Let $\nabla^{TM}$ be the Levi-Civita connection of
 $g^{TM}$ and $\nabla^F$ (resp. $\nabla^{F^\perp})$ be the
 restriction of $\nabla^{TM}$ to $F$ (resp. $F^\perp$). That is,
 $$\nabla^F=P\nabla^{TM}P,~~\nabla^{F^\perp}=P^\perp\nabla^{TM}P^\perp.\eqno(2.2)$$
We assume that $F$ is oriented, spin and carries a fixed spin
structure. We also assume that $F^\perp$ is oriented and that both
$2p={\rm dim}F$ and $q={\rm dim}F^{\perp}$ are even.\\
  \indent Let $S(F)$ be the bundle of spinors associated to
  $(F,g^F)$. For any $X\in \Gamma(F),$ denote by $c(X)$ the Clifford
  action of $X$ on $S(F)$. Since ${\rm dim}F$ is even, we have the
  splitting $S(F)=S_+(F)\oplus S_-(F)$ and $c(X)$ exchanges
  $S_+(F)$ and $S_-(F)$.\\
  \indent Let $\wedge(F^{\perp,\star})$ be the exterior algebra bundle
  of $F^{\perp}$. Then $\wedge(F^{\perp,\star})$ carries a
  canonically induced metric $g^{\wedge(F^{\perp,\star})}$ from
  $g^{F^\perp}$. For any $U\in \Gamma(F^\perp)$, let $U^*\in \Gamma(F^{\perp,*})$
  be the corresponding dual of $U$ with respect to $g^{F^\perp}$.
  Now for $U\in \Gamma(F^\perp)$, set
 $$c(U)=U^*\wedge-i_U,~~\widehat{c}(U)=U^*\wedge+i_U,\eqno(2.3)$$
 where $U^*\wedge$ and $i_U$ are the exterior and inner
 multiplication. Let $h_1.\cdots,h_q$ be an oriented local
 orthonormal basis of $F^\perp$. Then
 $\tau=(-\sqrt{-1})^{\frac{q(q+1)}{2}}c(h_1)\cdots c(h_q)$ and $\tau
 ^2=1$. Now the $+1$ and $-1$ eigenspaces of $\tau$ give a splitting
  $\wedge(F^{\perp,\star})=\wedge_+(F^{\perp,\star})\oplus
  \wedge_-(F^{\perp,\star}).$ Let
  $S(F)\widehat{\otimes}\wedge(F^{\perp,\star})$ be the ${\bf Z}_2$
  graded tensor product of $S(F)$ and $\wedge(F^{\perp,\star})$. For
$X\in \Gamma(F),~U\in \Gamma(F^\perp)$, the operators
$c(X),~c(U),~\widehat{c}(U)$ extend naturally to
$S(F)\widehat{\otimes}\wedge(F^{\perp,\star})$ and they are
anticommute. The connections $\nabla^F,~\nabla^{F^\perp}$ lift to
$S(F)$ and $\wedge(F^{\perp,\star})$ naturally. We write them
$\nabla^{S(F)}$ and $\nabla^{\wedge(F^{\perp,\star})}$. Then
$S(F)\widehat{\otimes}\wedge(F^{\perp,\star})$ carries the induced
tensor product connection
$\nabla^{S(F)\widehat{\otimes}\wedge(F^{\perp,\star})}$.\\
 \indent Let $S\in \Omega(T^*M)\otimes \Gamma({\rm End}(TM))$ be
 defined by
 $$\nabla^{TM}=\nabla^{F}+\nabla^{F^\perp}+S.\eqno(2.4)$$
 Then for any $X\in \Gamma(TM)$, $S(X)$ exchanges $\Gamma(F)$ and
 $\Gamma(F^\perp)$ and is skew-adjoint with respect to $g^{TM}$.
 Let $V$ be a complex vector bundle with the metric connection
 $\nabla^V$. Then
$S(F)\widehat{\otimes}\wedge(F^{\perp,\star})\otimes V$ carries the
induced tensor product connection
$\nabla^{S(F)\widehat{\otimes}\wedge(F^{\perp,\star})\otimes V}$.
 Let $\{f_i\}_{i=1}^{2p}$ be an oriented orthonormal basis of $F$.
 Let$$\widetilde{\nabla}=\nabla^{S(F)\widehat{\otimes}\wedge(F^{\perp,\star})}+
\frac{1}{2}\sum_{j=1}^{2p}\sum_{s=1}^q<S(.)f_j,h_s>c(f_j)c(h_s)$$
 $$\widetilde{\nabla}^{F,V}=\widetilde{\nabla}
 \otimes {\rm Id}_V+{\rm Id}_{S(F)\widehat{\otimes}\wedge(F^{\perp,\star})}\otimes
 \nabla^V.\eqno(2.5)$$
 \noindent Since the vector bundle $F^\perp$ might well be non-spin, Liu and
 Zhang [LZ] introduced the following sub-Dirac operator:\\

\noindent{\bf Definition 2.1} Let $D_{F,V}$ be the operator mapping
from $\Gamma({S(F)\widehat{\otimes}\wedge(F^{\perp,\star})\otimes
V})$ to itself defined by
$$D_{F,V}=\sum_{i=1}^{2p}c(f_i)\widetilde{\nabla}^{F,V}_{f_i}
+\sum_{s=1}^{q}c(h_s)\widetilde{\nabla}^{F,V}_{h_s}.\eqno(2.6)$$

\indent Let $\triangle^{F,V}$ be the Bochner Laplacian defined by
$$\triangle^{F,V}:=-\sum_{i=1}^{2p}(\widetilde{\nabla}^{F,V}_{f_i})^2
-\sum_{s=1}^{q}(\widetilde{\nabla}^{F,V}_{h_s})^2+
\widetilde{\nabla}^{F,V}_{\sum_{i=1}^{2p}\nabla^{TM}_{f_i}f_i} +
\widetilde{\nabla}^{F,V}_{\sum_{s=1}^{q}\nabla^{TM}_{h_s}h_s}.
\eqno(2.7)$$
 Let $r_M$ be the scalar curvature of the metric $g^{TM}$. Let
 $R^{F^\bot}$ and $R^V$ be curvature of $F^\bot$ and $V$. Then we
 have the following Lichnerowicz formula for $D_{F,V}$.\\

 \noindent {\bf Theorem 2.2([LZ])} {\it The following identity
 holds}
 $$D^2_{F,V}=\triangle^{F,V}+\frac{1}{2}\sum_{i,j=1}^{2p}c(f_i)c(f_j)R^V(f_i,f_j)$$
$$+\sum_{i=1}^{2p}\sum_{s=1}^qc(f_i)c(h_s)R^V(f_i,h_s)
+\frac{1}{2}\sum_{s,t=1}^{q}c(h_s)c(h_t)R^V(h_s,h_t)$$
$$+\frac{r_M}{4}+\frac{1}{4}\sum_{i=1}^{2p}\sum_{r,s,t=1}^q\left<R^{F^\bot}(f_i,h_r)h_t,h_s\right>
c(f_i)c(h_r)\widehat{c}(h_s)\widehat{c}(h_t)$$
$$+\frac{1}{8}\sum_{i,j=1}^{2p}\sum_{s,t=1}^q\left<R^{F^\bot}(f_i,f_j)h_t,h_s\right>
c(f_i)c(f_j)\widehat{c}(h_s)\widehat{c}(h_t)$$
$$+\frac{1}{8}\sum_{s,t,r,l=1}^q\left<R^{F^\bot}(h_r,h_l)h_t,h_s\right>
c(h_r)c(h_l)\widehat{c}(h_s)\widehat{c}(h_t).\eqno(2.8)$$\\

 \indent When $V$ is a complex line bundle, we write $D_{F}$ instead
of $D_{F,E}$. For the sub-Dirac operator $D_{F}$ we will calculate
the bosonic part of the spectral action. It is defined to be the
number of eigenvalues of $D_{F}$ in the interval $[-\wedge,\wedge]$
with $\wedge\in {\bf R}^+$. As in [CC1], it is expressed as
$$ I={\rm tr}\widehat{F}\left(\frac{D^2_F}{\wedge^2}\right).$$
\noindent Here tr denotes the operator trace in the $L^2$ completion
of $\Gamma (S(F)\widehat{\otimes}\wedge(F^{\perp,\star}))$, and
$\widehat{F}:{\bf R}^+\rightarrow {\bf R}^+$ is a cut-off function
with support in the interval $[0,1]$ which is constant near the
origin. Let ${\rm dim}M=m$. By Theorem 2.2, we have the heat trace
asymptotics for $t\rightarrow 0$,
$${\rm tr}(e^{-tD_F^2})\sim \sum_{n\geq
0}t^{n-\frac{m}{2}}a_{2n}(D_F^2).$$ One uses the Seely-deWitt
coefficients $a_{2n}(D_F^2)$ and $t=\wedge^{-2}$ to obtain an
asymptotics for the spectral action when ${\rm dim} M=4$ [CC1]
$$I={\rm tr}\widehat{F}\left(\frac{D^2_F}{\wedge^2}\right)\sim
\wedge^4F_4a_0(D^2_F)+\wedge^2F_2a_2(D^2_F)+\wedge^0F_0a_4(D^2_F)~~{\rm
as} ~~\wedge\rightarrow \infty \eqno(2.9)$$ \noindent with the first
three moments of the cut-off function which are given by
$F_4=\int_0^{\infty}s\widehat{F}(s)ds,$\\
$F_2=\int_0^{\infty}\widehat{F}(s)ds$
and $F_0=\widehat{F}(0)$. Let
$$-E=\frac{r_M}{4}+W=\frac{r_M}{4}+\frac{1}{4}\sum_{i=1}^{2p}\sum_{r,s,t=1}^q\left<R^{F^\bot}(f_i,h_r)h_t,h_s\right>
c(f_i)c(h_r)\widehat{c}(h_s)\widehat{c}(h_t)$$
$$+\frac{1}{8}\sum_{i,j=1}^{2p}\sum_{s,t=1}^q\left<R^{F^\bot}(f_i,f_j)h_t,h_s\right>
c(f_i)c(f_j)\widehat{c}(h_s)\widehat{c}(h_t)$$
$$+\frac{1}{8}\sum_{s,t,r,l=1}^q\left<R^{F^\bot}(h_r,h_l)h_t,h_s\right>
c(h_r)c(h_l)\widehat{c}(h_s)\widehat{c}(h_t),\eqno(2.10)$$ and
$$\Omega_{ij}=\widetilde{\nabla}_{e_i}\widetilde{\nabla}_{e_j}-\widetilde{\nabla}_{e_j}\widetilde{\nabla}_{e_i}
-\widetilde{\nabla}_{[e_i,e_j]},\eqno(2.11)$$ where $e_i$ is $f_i$
or $h_s$. We use [G, Thm 4.1.6] to obtain the first three
coefficients of the heat trace asymptotics:
$$a_0(D_F)=(4\pi)^{-\frac{m}{2}}\int_M{\rm tr}(\rm
Id)dvol,\eqno(2.12)$$
$$a_2(D_F)=(4\pi)^{-\frac{m}{2}}\int_M{\rm
tr}[(r_M+6E)/6]dvol,\eqno(2.13)$$
$$a_4(D_F)=\frac{(4\pi)^{-\frac{m}{2}}}{360}\int_M{\rm
tr}[-12R_{ijij,kk}+5R_{ijij}R_{klkl}$$
$$-2R_{ijik}R_{ljlk}+2R_{ijkl}R_{ijkl}-60R_{ijij}E+180E^2+60E_{,kk}+30\Omega_{ij}
\Omega_{ij}]dvol.\eqno(2.14)$$

Since ${\rm
dim}[S(F)\widehat{\otimes}\wedge(F^{\perp,\star})]=2^{p+q}$ and
$m=2p+q$, then we have
$a_0(D_F)=\frac{1}{2^p\pi^{p+\frac{q}{2}}}\int_Mdvol.$ By Clifford
relations and cyclicity of the trace and the trace of the odd degree
operator being zero, we have
$${\rm tr}(c(f_i))=0;~{\rm tr}(c(f_i)c(f_j))=0 ~{\rm for} ~i\neq j;$$
$$~{\rm tr}(c(h_r)c(h_l)\widehat{c}(h_s)\widehat{c}(h_t))=0, ~{\rm for}
~r\neq l.\eqno(2.15)$$ and
$${\rm
tr}E=-2^{p+q}\cdot\frac{r_M}{4},~~a_2(D_F)=-\frac{1}{12\cdot2^p\pi^{p+\frac{q}{2}}}\int_Mr_Mdvol.\eqno(2.16)$$
Let $I_1,I_2, I_3$ denote respectively the last three terms in
(2.10). By (2.15), we have $${\rm tr}(E^2)={\rm
tr}(\frac{r_M^2}{16}+W^2)={\rm
tr}(\frac{r_M^2}{16}+I_1^2+I_2^2+I_3^2).\eqno(2.17)$$
 $${\rm
tr}(I_1^2)=\frac{1}{16}\sum_{i,i'=1}^{2p}\sum_{r,r',s,s',t,t'=1}^q\left<R^{F^\bot}(f_i,h_r)h_t,h_s\right>
\left<R^{F^\bot}(f_{i'},h_{r'})h_{t'},h_{s'}\right>$$ $$\cdot{\rm
tr}[c(f_i)c(h_r)\widehat{c}(h_s)\widehat{c}(h_t)
c(f_{i'})c(h_{r'})\widehat{c}(h_{s'})\widehat{c}(h_{t'})]\eqno(2.18)$$
Similar to (2.15), we have $${\rm
tr}[c(f_i)c(h_r)\widehat{c}(h_s)\widehat{c}(h_t)
c(f_{i'})c(h_{r'})\widehat{c}(h_{s'})\widehat{c}(h_{t'})]$$
$$=-\delta_i^{i'}\delta_r^{r'}2^p{\rm
tr}_{\wedge(F^{\perp,\star})}[\widehat{c}(h_s)\widehat{c}(h_t)\widehat{c}(h_{s'})\widehat{c}(h_{t'})]\eqno(2.19)$$
Since $t\neq s,~t'\neq s'$. we get $$ {\rm
tr}_{\wedge(F^{\perp,\star})}[\widehat{c}(h_s)\widehat{c}(h_t)\widehat{c}(h_{s'})\widehat{c}(h_{t'})]
=(\delta_t^{s'}\delta_s^{t'}-\delta_t^{t'}\delta_s^{s'})2^q\eqno(2.20)$$
By (2.19) and (2.20), we have
$${\rm
tr}(I_1^2)=\frac{2^{p+q}}{8}\sum_{i=1}^{2p}\sum_{r,s,t=1}^q\left<R^{F^\bot}(f_i,h_r)h_t,h_s\right>^2.\eqno(2.21)$$
Similarly we have
$${\rm
tr}(I_2^2)=\frac{2^{p+q}}{16}\sum_{i,j=1}^{2p}\sum_{s,t=1}^q\left<R^{F^\bot}(f_i,f_j)h_t,h_s\right>^2;\eqno(2.22)$$
$${\rm
tr}(I_3^2)=\frac{2^{p+q}}{16}
\sum_{s,t,r,l=1}^q\left<R^{F^\bot}(h_r,h_l)h_t,h_s\right>^2.\eqno(2.23)$$
So we get
$${\rm tr
E^2}=\frac{2^{p+q}}{16}r_M^2+\frac{2^{p+q}}{16}||R^{F^\bot}||^2,\eqno(2.24)$$
where
$$||R^{F^\bot}||^2=2\sum_{i=1}^{2p}\sum_{r,s,t=1}^q\left<R^{F^\bot}(f_i,h_r)h_t,h_s\right>^2$$
$$+\sum_{i,j=1}^{2p}\sum_{s,t=1}^q\left<R^{F^\bot}(f_i,f_j)h_t,h_s\right>^2+
\sum_{s,t,r,l=1}^q\left<R^{F^\bot}(h_r,h_l)h_t,h_s\right>^2.\eqno(2.25)$$
Nextly we compute ${\rm tr}[\Omega_{ij} \Omega_{ij}]$ in a local
coordinate, so we can assume that $M$ is spin and
$\widetilde{\nabla}$ is the standard twisted connection on the
twisted spinors bundle $S(TM)\otimes S(F^\bot)$. Then
$$\Omega_{ij}=R^{S(TM)}(e_i,e_j)\otimes {\rm Id}_{S(F^\bot)}+{\rm
Id}_{S(TM)}\otimes R^{S(F^\bot)}(e_i,e_j)$$
$$=-\frac{1}{4}R^M_{ijkl}c(e_k)c(e_l)
\otimes {\rm Id}_{S(F^\bot)}-\frac{1}{4}{\rm Id}_{S(TM)}\otimes
\left<R^{F^\bot}(e_i,e_j)h_s,h_t\right>c(h_s)c(h_t).\eqno(2.26)$$
Similar to the computations of ${\rm tr E^2}$, we get
$${\rm
tr}[\Omega_{ij}
\Omega_{ij}]=-\frac{2^{p+q}}{8}(R_{ijkl}^2+||R^{F^\bot}||^2)\eqno(2.27)$$
By the divergence theorem and (2.24) and (2.27), we have
$$a_4(D_F^2)=\frac{1}{360\cdot2^p\pi^{p+\frac{q}{2}}}\int_M\left(\frac{5}{4}r_M^2-2R_{ijik}R_{ljlk}-\frac{7}{4}
R_{ijkl}^2+\frac{15}{2}||R^{F^\bot}||^2\right)dvol.\eqno(2.28)$$

\section{The spectral
      action  for the Standard Model associated to sub-Dirac
      operators}
       \quad In this section, we let $m=4$. We consider the product space $\cal{H}$
 of the $L^2$ completion
of $\Gamma (S(F)\widehat{\otimes}\wedge(F^{\perp,\star}))$ and a
finite dimensional Hilbert space $ {\cal H}_f$ (called internal
Hilbert space). The specific particle model is encoded in $ {\cal
H}_f$. On the bundle
$S(F)\widehat{\otimes}\wedge(F^{\perp,\star})\otimes{\cal H}_f$ one
considers a connection $\widetilde{\nabla}^{F,{\cal H}_f}$ in (2.5)
and $\nabla^{{\cal H}_f}$ is a covariant derivative in the trivial
bundle ${{\cal H}_f}$ induced gauge fields. The associated Dirac
operator to $\widetilde{\nabla}^{F,{\cal H}_f}$ is called $D^f_F$.
The generalized Dirac operator of the Standard Model $D_{F,\Phi}$
contains the Higgs boson, Yukawa couplings, neutrino masses and the
CKM-matrix encoded in a field $\Phi$ of endomorphisms of ${{\cal
H}_f}$. We define $D_{F,\Phi}$ for sections $\psi\otimes \chi\in
{\cal H}$ as
$$D_{F,\Phi}(\psi\otimes
\chi)=D^f_F(\psi\otimes
\chi)+\gamma_5\psi\otimes\Phi\chi,\eqno(3.1)$$ where
$\gamma_5=e_0e_1e_2e_3$ is the volume element. We choose the same
$\Phi$ as $\Phi$ in [CC1]. The bosonic part of the Lagrangian of the
Standard Model is obtained by replacing $D_F$ by $D_{F,\Phi}$ in
(2.9). In (2.8), we write $D^2_{F,{\cal H}_f}=\triangle^{F,{\cal
H}_f}+W_1$. Then direct computations show
$$D_{F,\Phi}^2=\triangle^{F,{\cal H}_f}-E_\Phi,\eqno(3.2)$$
where the potential is defined as
$$E_\Phi(\psi\otimes
\chi)=-W_1(\psi\otimes
\chi)+\sum_{i=1}^4\gamma_5c(e_i)\cdot\psi\otimes[\nabla_{e_i}^{H_f},\Phi]\chi-\psi\otimes\Phi^2\chi.\eqno(3.3)$$
We denote the trace on ${\cal H}$ and on ${\cal H}_f$ as Tr and
${\rm tr}_f$. From (3.3), we have
$${\rm Tr}(E_\Phi)={\rm dim}{\cal
H}_f\cdot 2^{p+q-2}r_M-2^{p+q}{\rm tr}_f(\Phi^2).\eqno(3.4)$$\indent
For Seely-deWitt coefficient $a_4(D_{F,\Phi}^2)$ we also need to
calculate
$$(E_\Phi)^2(\psi\otimes
\chi)=W_1^2(\psi\otimes
\chi)+\sum_{i,j=1}^4\gamma_5c(e_i)\gamma_5c(e_j)\cdot\psi\otimes
[\nabla_{e_i}^{H_f},\Phi][\nabla_{e_j}^{H_f},\Phi]\chi$$
$$+\psi\otimes\Phi^4\chi-2E\psi\otimes\Phi^2\chi+
\frac{1}{2}\sum_{i,j=1}^{2p}c(f_i)c(f_j)\psi\otimes [\Phi^2R^{{\cal
H}_f}(f_i,f_j)+R^{{\cal H}_f}(f_i,f_j)\Phi^2]\chi$$
$$+\sum_{i=1}^{2p}\sum_{s=1}^qc(f_i)c(h_s)\psi\otimes [\Phi^2R^{{\cal H}_f}(f_i,h_s)
+R^{{\cal H}_f}(f_i,h_s)\Phi^2]\chi$$
$$+\frac{1}{2}\sum_{s,t=1}^{q}c(h_s)c(h_t)\psi\otimes [\Phi^2R^{{\cal
H}_f}(h_s,h_t)+R^{{\cal H}_f}(h_s,h_t)\Phi^2]\chi$$
$$-\sum_{i=1}^4\gamma_5c(e_i)\psi\otimes(\Phi^2[\nabla_{e_i}^{H_f},\Phi]+[\nabla_{e_i}^{H_f},\Phi]\Phi^2)\chi
$$
$$+\sum_{i=1}^4(E\gamma_5c(e_i)\psi+\gamma_5c(e_i)E\psi)\otimes
[\nabla_{e_i}^{H_f},\Phi]\chi$$
$$-\frac{1}{2}\sum_{i,j,k=1}^4\gamma_5c(e_i)c(e_j)c(e_k)\psi\otimes[\nabla_{e_i}^{H_f},\Phi]
R^{{\cal H}_f}(e_j,e_k)\chi$$ $$-
\frac{1}{2}\sum_{i,j,k=1}^4c(e_j)c(e_k)\gamma_5c(e_i)\psi\otimes
R^{{\cal H}_f}(e_j,e_k)[\nabla_{e_i}^{H_f},\Phi] \chi.\eqno(3.5)$$
By Clifford relations and cyclicity of the trace and the trace of
the odd degree operator being zero, only the first four summands on
the right-hand side contribute to the trace of $(E_\Phi)^2$. By
direct computations, we get
$${\rm Tr}(E_\Phi^2)={\rm dim}{\cal
H}_f\frac{2^{p+q}}{16}(r_M^2+||R^{F^\bot}||^2)-2^{p+q-1}\sum_{i,j=1}^4{\rm
tr}_f(\Omega^f_{ij}\Omega^f_{ij})$$ $$+2^{p+q-1}r_M{\rm
tr}_f(\Phi^2)+2^{p+q}{\rm tr}_f(\Phi^4)+2^{p+q}\sum_{i=1}^4{\rm
tr}_f([\nabla_{e_i}^{H_f},\Phi]^2).\eqno(3.6)$$ By (2.27), we have
$${\rm Tr}(\widetilde{\Omega}^f_{ij}\widetilde{\Omega}^f_{ij})
=-{\rm dim}{\cal H}_f\cdot
\frac{2^{p+q}}{8}(R_{ijkl}^2+||R^{F^\bot}||^2)+2^{p+q}{\rm
tr}_f(\Omega^f_{ij}\Omega^f_{ij}).\eqno(3.7)$$
 We choose the finite space ${\cal H}_f$ according to the
 construction of the noncommutative Standard Model [CC1], ${\rm dim}{\cal
 H}_f=96$ and $\nabla^{{\cal H}_f}$ is the appropriate covariant
 derivative associated to the Standard Model gauge group
 $U(1)_Y\times SU(2)_\omega\times SU(3)_c.$
 We know that (for related notations see [HPS], [IKS]),
 $${\rm
tr}_f(\Omega^f_{ij}\Omega^f_{ij})=\frac{48}{5}g_3^2||G||^2+\frac{48}{5}g_2^2||F_1||^2+16g_1^2||B||^2,\eqno(3.8)$$
$${\rm
tr}_f([\nabla_{e_i}^{H_f},\Phi]^2)=4a|D_\nu\varphi|^2,~~{\rm
tr}_f(\Phi^2)=4a|\phi|^2+2c,~~{\rm
tr}_f(\Phi^4)=4b|\phi|^4+8e|\phi|^2+2d.\eqno(3.9)$$ Then we get
$$a_0(D_{F,\Phi})=\frac{96}{2^p\pi^{p+\frac{q}{2}}}\int_Mdvol,\eqno(3.10)$$
$$a_2(D_{F,\Phi})=\frac{1}{2^p\pi^{p+\frac{q}{2}}}\int_M(40r_M-4a|\phi|^2-2c)dvol,\eqno(3.11)$$
$$a_4(D_{F,\Phi})=\frac{1}{360\cdot 2^p\pi^{p+\frac{q}{2}}}\int_M
\left\{4000r_M^2-192R_{ijik}R_{ljlk}-168R^2_{ijkl}+120ar_M|\varphi|^2\right.$$
$$+60cr_M+720||R^{F^\bot}||^2-576g_3^2||G||^2
-576g_2^2||F_1||^2-960g_1^2||B||^2$$
$$\left.+720b|\varphi|^4+1440e|\varphi|^2+360d+720|D_\nu\varphi|^2\right\}dvol.\eqno(3.12)dvol$$
\indent In presence of the Standard Model fields we obtain
essentially one new term (apart from the usual suspects)
$$I_{\rm new}=\frac{2}{
2^p\pi^{p+\frac{q}{2}}}\int_M||R^{F^\bot}||^2dvol.\eqno(3.13)$$

\section{ The spectral action for foliations with boundary}

     \quad In this section, we let $M$ be a foliation with boundary
     $\partial M$. Let $\psi\in \Gamma(S(F)\widehat{\otimes}\wedge(F^{\perp,\star})$,
      We impose the Dirichlet boundary conditions $\psi|_{\partial
      M}=0$. With the Dirichlet boundary conditions, we have the heat trace
asymptotics for $t\rightarrow 0$ [BG],
$${\rm tr}(e^{-tD_F^2})\sim \sum_{n\geq
0}t^{\frac{n-m}{2}}a_{n}(D_F^2).$$ One uses the Seely-deWitt
coefficients $a_{n}(D_F^2)$ and $t=\wedge^{-2}$ to obtain an
asymptotics for the spectral action when ${\rm dim} M=4$ [ILV (18)]
$$I={\rm tr}\widehat{F}\left(\frac{D^2_F}{\wedge^2}\right)\sim
\wedge^4F_4a_0(D^2_F)+\wedge^3F_3a_1(D^2_F)$$
$$+\wedge^2F_2a_2(D^2_F)+\wedge
F_1a_3(D^2_F)+\wedge^0F_0a_4(D^2_F)~~{\rm as} ~~\wedge\rightarrow
\infty \eqno(4.1)$$ where
$F_k:=\frac{1}{\Gamma(\frac{k}{2})}\int_0^{\infty}\widehat{F}(s)s^{\frac{k}{2}-1}ds.$
Let $N=e_m$ be the inward pointing unit normal vector on $\partial
M$ and $e_i, 1\leq i\leq m-1$ be the orthonormal frame on
$T(\partial M)$. Let $ L_{ab}=(\nabla_{e_a}e_b,N)$ be the second
fundamental form and indices $\{a,b,\cdots\}$ range from $1$ through
$m-1$. We use [BG, Thm 1.1] to obtain the first five coefficients of
the heat trace asymptotics:
$$a_0(D_F)=(4\pi)^{-\frac{m}{2}}\int_M{\rm tr}(\rm
Id)dvol_M,\eqno(4.2)$$
$$a_1(D_F)=-4^{-1}(4\pi)^{-\frac{(m-1)}{2}}\int_{\partial M}{\rm tr}(\rm
Id)dvol_{\partial M},\eqno(4.3)$$
$$a_2(D_F)=(4\pi)^{-\frac{m}{2}}6^{-1}\{\int_M{\rm
tr}(r_M+6E)dvol_M+2\int_{\partial M}{\rm tr}(L_{aa})dvol_{\partial
M}\},\eqno(4.4)$$
$$a_3(D_F)=-4^{-1}(4\pi)^{-\frac{(m-1)}{2}}96^{-1}\{\int_{\partial M}{\rm
tr}(96E+16r_M$$
$$+8R_{aNaN}+7L_{aa}L_{bb}-10L_{ab}L_{ab})dvol_{\partial
M})\},\eqno(4.5)$$
$$a_4(D_F)=\frac{(4\pi)^{-\frac{m}{2}}}{360}\{
\int_M{\rm
tr}[-12R_{ijij,kk}+5R_{ijij}R_{klkl}$$
$$-2R_{ijik}R_{ljlk}+2R_{ijkl}R_{ijkl}-60R_{ijij}E+180E^2+60E_{,kk}+30\Omega_{ij}
\Omega_{ij}]dvol_M $$ $$+\int_{\partial M}{\rm
tr}(-120E;_N-18r_M;_N+120EL_{aa}+20r_ML_{aa}+4R_{aNaN}L_{bb}-12R_{aNbN}L_{ab}$$
$$+4R_{abcd}L_{ac}+24L_{aa;bb}
+40/21L_{aa}L_{bb}L_{cc}-88/7L_{ab}L_{ab}L_{cc}+320/21L_{ab}L_{bc}L_{ac})dvol_{\partial
M}\}.\eqno(4.6)$$ By (2.16) and (2.28) and the divergence theorem
for manifolds with boundary, we get
$$a_0(D_F)=\frac{1}{2^p\pi^{p+\frac{q}{2}}}\int_Mdvol_M,\eqno(4.7)$$
$$a_1(D_F)=-4^{-1}(4\pi)^{-\frac{(m-1)}{2}}2^{p+q}\int_{\partial M}dvol_{\partial M},\eqno(4.8)$$
$$a_2(D_F)=\frac{1}{12\cdot
2^p\pi^{p+\frac{q}{2}}}(-\int_Mr_Mdvol_M+4\int_{\partial
M}L_{aa}dvol_{\partial M}),\eqno(4.9)$$
$$a_3(D_F)=-4^{-1}(4\pi)^{-\frac{(m-1)}{2}}96^{-1}2^{p+q}\{\int_{\partial
M}(-8r_M$$ $$+8R_{aNaN}+7L_{aa}L_{bb}-10L_{ab}L_{ab})dvol_{\partial
M}\},\eqno(4.10)$$
$$a_4(D_F)=\frac{(4\pi)^{-\frac{m}{2}}}{360}2^{p+q}\{
\int_M\left(\frac{5}{4}r_M^2-2R_{ijik}R_{ljlk}-\frac{7}{4}
R_{ijkl}^2+\frac{15}{2}||R^{F^\bot}||^2\right)dvol_M$$ $$+
\int_{\partial M}{\rm
tr}(-51r_{M;N}-10r_ML_{aa}+4R_{aNaN}L_{bb}-12R_{aNbN}L_{ab}$$
$$+4R_{abcd}L_{ac}+24L_{aa;bb}
+40/21L_{aa}L_{bb}L_{cc}-88/7L_{ab}L_{ab}L_{cc}+320/21L_{ab}L_{bc}L_{ac})dvol_{\partial
M}\}.\eqno(4.11)$$\\

 \noindent {\bf Acknowledgement.} This work
was supported by NSFC No.10801027 and Fok Ying Tong Education
Foundation No. 121003.\\

\noindent{\large \bf References}\\

\noindent[BG] Branson, T. P.; Gilkey, P. B. The asymptotics
  of the Laplacian on a manifold with boundary,
   Comm. Partial Differential Equations 15 (1990), no. 2,
   245-272.\\
\noindent [CC1] Chamseddine, A. H.; Connes, A., The spectral action
principle. Comm. Math. Phys. 186 (1997), no. 3, 731-750.
\\
 \noindent[CC2] Chamseddine, A. H.; Connes, A.
Noncommutative geometric spaces with boundary: spectral action. J.
Geom. Phys. 61
(2011), no. 1, 317-332.\\
 \noindent[Co] Connes, A., Gravity coupled
 with matter and the foundation of non-commutative geometry. Comm. Math. Phys. 182 (1996), no. 1,
 155-176.\\
\noindent[EILS] Essouabri, D.; Iochum, B.; Levy, C.; Sitarz, A.,
 Spectral action on noncommutative torus. J. Noncommut. Geom. 2 (2008), no. 1,
 53-123.\\
\noindent[Gi]Gilkey, P. B., {\it Invariance theory, the heat
equation, and the Atiyah-Singer index theorem.} Mathematics Lecture
Series, 11. Publish or Perish, 1984 \\
\noindent [HPS]Hanisch, F.; Pf\"{a}ffle, F.; Stephan, C. A., The
spectral action for Dirac operators with skew-symmetric torsion.
Comm. Math. Phys. 300 (2010), no. 3, 877-888.\\
\noindent[IKS]Iochum, B.; Kastler, D.; Sch¨¹cker, T., On the
universal Chamseddine-Connes action. I. Details of the action
computation. J. Math. Phys. 38 (1997), no. 10, 4929-4950.\\
\noindent[ILS] Iochum, B.; Levy, C.; Sitarz, A.,
  Spectral action on ${\rm SU}_q(2)$. Comm. Math. Phys. 289 (2009), no. 1,
  107-155.\\
\noindent[ILV]Iochum, B.; Levy, C.; Vassilevich, D., Spectral action
for torsion with and without boundaries, arXiv:1008.3630.\\
\noindent [LZ] Liu, K.; Zhang, W., Adiabatic limits and foliations,
 Contemp. Math.,2001,195-208.\\

 \indent{  School of Mathematics and Statistics,
Northeast Normal University, Changchun Jilin, 130024, China }\\
\indent E-mail: {\it wangy581@nenu.edu.cn}\\

\end{document}